# On the Spin Period of the Neutron Star in the Ultraluminous X-Ray Source M51 ULX-8

Mehmet Hakan Erkut[1*]

[1*] Istanbul Bilgi University, Faculty of Engineering and Natural Sciences, 34060, İstanbul, Turkey, (ORCID: 0000-0003-1054-264X), mherkut@gmail.com

## Abstract

The recent discovery of periodic pulsations from several members of the ultraluminous X-ray source (ULX) family in nearby galaxies as well as in our own galaxy unveiled the nature of the accreting compact object. Neutron stars rather than black holes are currently believed to power a substantial number of ULXs whether or not pulsations are observed. The detection of cyclotron absorption lines in the X-ray spectrum of a ULX provides an alternative way to identify the compact object as a neutron star. Among the non-pulsating ULXs, the presence of a cyclotron resonance scattering feature (CRSF) in the spectrum of M51 ULX-8 has been reported. In the present work, the magnetic field strength on the surface of the neutron star in M51 ULX-8 is inferred from the energy of the observed CRSF to estimate the beaming fraction in X-ray emission and more importantly the observable range for the elusive neutron-star spin period to be hopefully discovered by the forthcoming space missions in the near future.

## Aşırı Parlak X-Işın Kaynağı M51 ULX-8'deki Nötron Yıldızının Dönme Periyodu Üzerine

## Öz

Galaksimizi de içermek üzere yakın galaksilerdeki aşırı parlak X-ışın kaynak (APX) ailesinin bazı üyelerinden geldiği kısa süre önce keşfedilen periyodik pulsasyonlar madde yığıştıran yoğun cismin doğasını ortaya çıkarmıştır. Pulsasyon gözlensin veya gözlenmesin, günümüzde, önemli sayıda APX'e karadeliklerden daha çok nötron yıldızlarının güç sağladığına inanılmaktadır. Bir APX'in X-ışın tayfındaki siklotron soğurma çizgilerinin saptanması yoğun cismin bir nötron yıldızı olarak tanımlanabilmesi için alternatif bir yol sağlamaktadır. Pulsasyon göstermeyen APX'lerin arasında bulunan M51 ULX-8'in tayfındaki bir siklotron rezonans saçılım yapısının (SRSY) varlığı rapor edilmişti. Bu çalışmada, gözlenen SRSY'nin enerjisinden M51 ULX-8'deki nötron yıldızının yüzeyindeki manyetik alan yeğinliği çıkarsanarak X-ışın salınımındaki hüzmeleme oranı ve daha da önemlisi yakın gelecekteki uzay görevleri sayesinde keşfedilmesi beklenen nötron yıldızının yakalanamamış dönme periyodunun gözlenebilme aralığı kestirilmektedir.

**Anahtar Kelimeler:** Nötron yıldızları, Yığışma, X-ışın çift yıldızları.

---

[*] Corresponding Author: mherkut@gmail.com

# 1. Introduction

Ultraluminous X-ray sources (ULXs) are the brightest objects in X-rays after active galactic nuclei (AGN) and supernovae observed in nearby galaxies. Unlike AGN, ULXs are point sources that do not generally coincide with the nucleus of the galaxy. The X-ray luminosities of ULXs are usually in the ~ $10^{39}$-$10^{41}$ erg s$^{-1}$ range and therefore exceed the Eddington luminosity of a stellar mass object such as a black hole of mass < 10 M$_\odot$ assuming that the X-ray emission is isotropic. It is now widely believed that the super-Eddington accretion of matter onto the stellar mass black holes and neutron stars in high-mass X-ray binaries and the geometrical beaming of radiation which measures the degree of anisotropy in emission can account for the standard $10^{39}$-$10^{41}$ erg s$^{-1}$ super-Eddington luminosity range. Albeit rare, however, the hyperluminous members of the ULX population with luminosities $\geq 10^{41}$ erg s$^{-1}$ are likely to be powered by the so-called accreting intermediate mass black holes of $10^2$-$10^5$ M$_\odot$ (Kaaret et al., 2017).

The recent discovery of periodic pulsations in X-rays from seven ULXs have revealed that the accreting compact objects in these sources are in fact neutron stars rather than black holes. The neutron-star spin periods in these ULXs which consist of one galactic (Swift J0243.6+6124) and six extragalactic sources (ULX NGC 5907, ULX NGC 7793 P13, M82 X-2, NGC 300 ULX1, M51 ULX-7, NGC 1313 X-2) have been measured in the 0.417-31.6 s range (Bachetti et al., 2014; Fürst et al., 2016; Israel et al., 2017; Carpano et al., 2018; Wilson-Hodge et al., 2018; Sathyaprakash et al., 2019; Rodríguez Castillo et al., 2020). Both the super-Eddington (supercritical) accretion with moderate beaming and the existence of sufficiently strong pulsar magnetic fields to raise the critical luminosity level via reduction of the scattering cross section of the electron have been invoked in order to explain the observed super-Eddington luminosities of such ultraluminous X-ray pulsars (Bachetti et al., 2014; Ekşi et al., 2015; Erkut et al., 2020).

The lack of pulsations from ULXs cannot be a direct evidence for black holes. The accreting compact object in a ULX might be a neutron star and yet the spin modulation of the pulsar emission may not be observed due to either an optically thick environment enshrouding the magnetosphere and thus smearing out the X-ray pulsations (Ekşi et al., 2015) or a supercritical propeller phase fed by the spin-down power while accretion is halted (Erkut et al., 2019). Sufficiently small beaming fractions might also account for the absence of pulsations from most ULXs (Erkut et al., 2020). Among pulsating and non-pulsating ULXs, two sources namely NGC 300 ULX1 and M51 ULX-8 exhibited cyclotron absorption lines in their spectra. In the spectrum of the non-pulsating source M51 ULX-8, the cyclotron line was detected at ~ 4.52 keV (Brightman et al., 2018; Middleton et al., 2019). Despite the failure to detect any pulsations, the accreting compact object in M51 ULX-8 could certainly be identified as a neutron star owing to this cyclotron resonance scattering feature (CRSF).

In this paper, the main focus is given to the surface magnetic field strength of the neutron star in M51 ULX-8, which is inferred from the line energy of the observed CRSF. Our aim is to estimate the elusive spin period of the neutron star using the allowed region in the parameter space of beaming fraction and spin period. Our results would be helpful for the search of periodic pulsations to be discovered in the near future.

# 2. Material and Method

## 2.1. Basic Equations and Assumptions

The maximum critical luminosity of an accreting star increases following the reduction of the scattering cross section of electrons due to increase in the stellar surface magnetic field (Paczynski, 1992). In the absence of any magnetic field, the maximum critical luminosity is determined by the Eddington limit, that is, $L_c = L_E$. In accordance with Erkut et al. (2020), the maximum critical luminosity of an accreting neutron star can be expressed as

$$L_c = \left[1 + 311\left(\frac{B}{B_c}\right)^{4/3}\right]L_E \qquad (1)$$

where $B$ is the neutron-star surface magnetic field, $B_c$ is the quantum critical magnetic field for electrons given by

$$B_c = \frac{m_e^2 c^3}{\hbar e} \cong 4.4 \times 10^{13}\,\text{G}, \qquad (2)$$

and $L_E$ is the Eddington luminosity that can be written as

$$L_E = 4\pi G M_* m_p c / \sigma_T. \qquad (3)$$

Here, $m_e$ and $m_p$ represent the electron and proton masses, respectively, $M_*$ is the neutron-star mass, $\sigma_T$ is the Thomson cross section of the electron, and $e$ is the elementary charge. All other symbols for fundamental constants have their usual meanings such as the speed of light, the gravitational constant, and the reduced Planck's constant.

The X-ray luminosity of the source is related to the X-ray flux through

$$L_X = 4\pi b d^2 F_X, \qquad (4)$$

where $b \leq 1$ is the beaming fraction of the X-ray emission, $d$ is the source distance, and $F_X$ is the unabsorbed X-ray source flux received by the observer (Erkut et al., 2020). Assuming that the accretion luminosity is mostly emitted in X-rays, the X-ray luminosity can be written as

$$L_X = G M_* \dot{M}_* / R_* = \varepsilon \dot{M}_* c^2. \qquad (5)$$

Here, $R_*$ is the radius of the neutron star, $\varepsilon$ is the efficiency of gravitational energy release, and $\dot{M}_*$ is the rate of mass accretion onto the neutron-star surface (Erkut et al., 2020).

In a high-mass X-ray binary, the super-Eddington rate of mass transfer from the mass-donor star to the accretion disk around the neutron star must satisfy

$$\dot{M}_0 > \dot{M}_E = \frac{L_E}{\varepsilon c^2} \qquad (6)$$

(Erkut et al., 2020). The mass accretion rate onto the neutron-star surface is only a fraction of this mass transfer rate, which is determined by the ratio of the inner disk radius $R_{in}$ to the so-called spherization radius $R_{sp}$ if $R_{in} < R_{sp}$. Otherwise, when $R_{in} >$ $R_{sp}$, both the mass transfer and mass accretion rates are the same (Shakura & Sunyaev, 1973).

For sources where the neutron stars are close to spin equilibrium, the lower limit of the beaming fraction can be written as

$$b_{min} \cong 0.13 \left(\frac{M_*}{1.4 M_{Sun}}\right)\left(\frac{10^{-11} \text{erg s}^{-1} \text{cm}^{-2}}{F_X}\right)\left(\frac{1 \text{Mpc}}{d}\right)^2 \quad (7)$$

using the super-Eddington mass-transfer-rate condition in Equation (6) for both $R_{in} < R_{sp}$ and $R_{in} > R_{sp}$ (Erkut et al., 2020).

The upper limit of the beaming fraction can be estimated using the subcritical luminosity condition $L_X \leq L_c$, where the maximum critical luminosity is given by Equation (1). Substituting Equation (4) for the X-ray luminosity of the source, it follows from $L_X \leq L_c$ that

$$b \leq \left[1 + 311\left(\frac{B}{B_c}\right)^{4/3}\right] \frac{L_E}{4\pi d^2 F_X}. \quad (8)$$

Here, Equation (8) yields the upper limit of $b$ provided the right hand side of the inequality is less than 1. Otherwise, $b = 1$ is the upper limit of the beaming fraction which can be expressed as

$$b_{max} = \min\left\{1, \left[1 + 311\left(\frac{B}{B_c}\right)^{4/3}\right] \frac{L_E}{4\pi d^2 F_X}\right\}. \quad (9)$$

Using the definition of the fastness parameter $\omega$ and the inner radius of the accretion disk around the neutron star in a ULX, the dipolar magnetic field strength on the surface of the neutron star has been obtained by Erkut et al. (2020) as in their Equation (16), which is employed here to solve for the beaming fraction:

$$b = \frac{\pi^{4/3} R_*^5 B^2 \delta}{2^{5/3} \omega^{7/3} (GM_*)^{2/3} F_X d^2 P^{7/3}}. \quad (10)$$

Note that the beaming fraction depends, in addition to other parameters, on the spin period of the neutron star, $P$, and the width of the boundary region, $\delta$, where the magnetic field lines of the neutron star couple with the innermost disk matter due to Rayleigh-Taylor and Kelvin-Helmholtz instabilities that are believed to play role in the magnetic reconnection and heating in the atmosphere of the Sun as well (Çavuş & Hoshoudy, 2019).

## 2.2. Magnetic Field Inferred from the Cyclotron Absorption Line

In the spectra of X-ray sources harboring strongly magnetized neutron stars, cyclotron features are usually detected in the form of absorption lines as a result of resonant scattering of photons by charged particles such as electrons and protons with quantized energies, also known as Landau levels, associated with the motion of particles perpendicular to the magnetic field. These features are therefore expected to be produced near the magnetic poles of accreting neutron stars when the outgoing radiation, which is emitted close to the surface of the neutron star, interacts with the infalling hot plasma of electrons and protons gyrating along the field lines.

The cyclotron line energies are given by the integer multiples of the energy quantum $\hbar\omega_{cyc}$, where

$$\omega_{cyc} = \frac{eB}{mc} \quad (11)$$

is called the gyro frequency of the charged particle of mass $m$. The observed energy of the cyclotron feature is redshifted if the line is generated near a gravitating object such as a neutron star. The energy of the cyclotron feature to be detected in the spectrum of a neutron-star source can therefore be expressed as

$$E_{cyc} = \frac{n\hbar e}{(1+z)mc} B. \quad (12)$$

Here, the gravitational redshift can be written as

$$z = \frac{1}{\sqrt{1 - \frac{2GM_*}{c^2 R_{cyc}}}} - 1, \quad (13)$$

where $R_{cyc}$ is the radius at which the photon is scattered. The fundamental line and its first harmonic correspond to the quantum numbers $n = 1$ (transition from the ground state to the first excited state) and $n = 2$ (transition from the ground state to the second excited state), respectively (Staubert et al., 2019).

Using $m = m_e$, the magnetic field can be estimated from Equation (12) as

$$B_e \cong \frac{E_{cyc}}{11.6 \text{ keV}} \left(\frac{1+z}{n}\right) \times 10^{12} \text{G} \quad (14)$$

for the electron cyclotron resonance scattering feature (eCRSF). Similarly, substituting $m_p$ for $m$ in Equation (12),

$$B_p \cong \frac{E_{cyc}}{6.32 \text{ keV}} \left(\frac{1+z}{n}\right) \times 10^{15} \text{G} \quad (15)$$

is found as the magnetic field estimate for the proton cyclotron resonance scattering feature (pCRSF).

## 3. Results and Discussion

### 3.1. Neutron-Star Magnetic Field in M51 ULX-8

Measuring the centroid energy of a CRSF in the source spectrum can provide a direct way of determining the magnetic field strength on the surface of a neutron star. The values estimated for $B$ may however differ from each other by orders of magnitude depending on the interpretation of the CRSF in terms of electrons or protons. In Figure 1, the run of the surface magnetic field of the neutron star in M51 ULX-8 is displayed as the compactness changes from 0.5 to 2, which corresponds to the widest possible range with regard to the majority of neutron-star equations of state (Erkut et al., 2016). The compactness is defined as the neutron-star mass to radius ratio, where the mass and radius are normalized to one solar mass and 10 km, respectively. The field values inferred from the eCRSF and pCRSF interpretations are calculated by evaluating the observed line energy $E_{cyc} \approx 4.52$ keV (Brightman et al., 2018) in Equations (14) and (15). The feature at 4.52 keV is considered as the fundamental cyclotron line with $n = 1$ in line with Brightman et al. (2018) and the line is assumed to be generated close to the surface of the neutron star while estimating the gravitational redshift (Equation 13), i.e., $R_{cyc} = R_*$ (Becker et al., 2012). Note that the maximum variation of $B$, for either electrons or protons, is limited within a factor of ~ 1.5 over the whole range of compactness and yet the field values inferred from protons are higher than those inferred from electrons by more than three orders of magnitude. The magnetic field values reckoned by pCRSFs indeed surpass the quantum critical magnetic field for electrons (Equation 2), where the electron's rest mass and cyclotron energies are comparable.

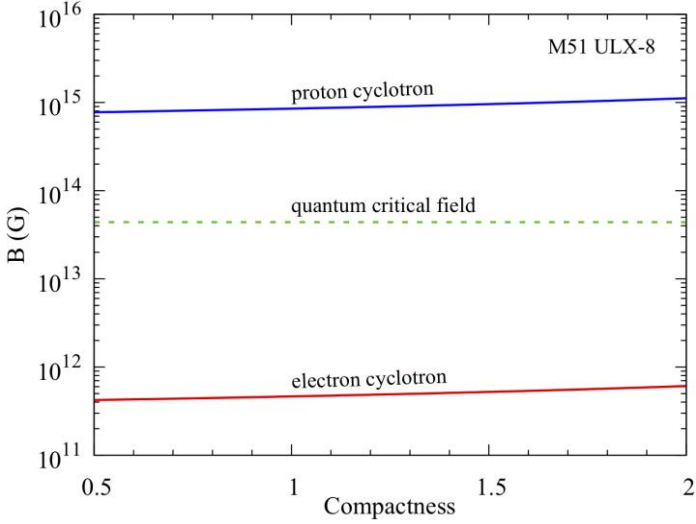

*Figure 1. Magnetic field estimate for the widest range of the neutron-star compactness (mass to radius ratio) using the CRSF detected at 4.52 keV in the X-ray spectrum of M51 ULX-8 with regard to electrons and protons.*

One of the ways to differentiate between eCRSF and pCRSF is to refer to the relative width of the observed absorption line. The CRSF detected in the 0.5-8 keV X-ray spectrum of M51 ULX-8 by Chandra Observatory (Brightman et al., 2018) was modeled with the use of a Gaussian absorption line. A Gaussian width of $\sigma \approx 0.11$ keV was inferred from the spectral fit to data. The relative width was therefore estimated as $\sigma/E_{cyc} \approx 0.02$, which is much smaller than the typical values of the relative widths of eCRSFs observed in Galactic neutron-star binaries, where $\sigma/E_{cyc} > 0.1$ in most cases. If the line widths are mainly determined by the thermal Doppler broadening caused by the thermal velocities of charged particles, electrons rather than protons are expected to produce broader lines as the former is much lighter than the latter. Based on the similarities between the relative widths of the line in M51 ULX-8 and the absorption lines observed in several soft gamma-ray repeaters, which are believed to possess magnetar fields (~$10^{14}$-$10^{15}$ G), Brightman et al. (2018) came to the conclusion that the proton cyclotron interpretation is more likely.

The time resolved analysis by Middleton et al. (2019) to decompose the spectrum of M51 ULX-8 into the multiple components for depicting the broad-band behavior of the source, however, brought an upper limit of ~ $10^{12}$ G on the dipole component of the magnetic field on the surface of the neutron star even though multipole fields of much higher strengths might still be plausible. A broader line of width $\sigma \approx 1$ keV also seemed to be allowed by one of their time-averaged spectral fits, thus favoring the electron cyclotron interpretation for the observed feature at ~ 4.5 keV.

In the presence of unresolved issues that are merely related to the classification of the firmly detected cyclotron line in M51 ULX-8 as due to electrons or protons, while ruling out other possibilities such as the formation of absorption line due to any known atomic transitions (Brightman et al., 2018), both cases, namely electron and proton cyclotron interpretations, will be considered equally probable in the search for the elusive spin period of the neutron star in the next subsection.

### 3.2. Neutron-Star Spin Period in M51 ULX-8

Most of the pulsating ULXs (PULXs) with measured spin period $P$ and period derivatives $\dot{P}$ are characterized by the spin-up timescales of $P/\dot{P} \sim 10^2$ yr that are so short, when compared to their lifetime ~$10^6$ yr, that the neutron stars in these systems are expected to be near spin equilibrium as a result of the interaction of their magnetospheres with their accretion disks (Bachetti et al., 2014; Fürst et al., 2016). If the neutron star is indeed very close to spin equilibrium, surface magnetic fields of magnetar strength might be needed to account for both the observed spin-up rate of the object and the super-Eddington luminosity in X-rays by increasing the maximum critical luminosity of the accreting neutron star via the reduction of the scattering cross section (Ekşi et al., 2015).

As suggested by Erkut et al. (2020), the neutron-star ULXs in high-mass X-ray binaries may never reach true spin equilibrium in the course of their lifetime as they might swing between propeller and accretion states with spin down and spin up, respectively and yet they may still be close to spin equilibrium with a fastness parameter in the neighborhood of its critical value, i.e., $\omega \approx \omega_c$.

The lack of pulsations from M51 ULX-8 might be due to the high optical thickness of the medium engulfing the pulsed emission from the neutron star (Ekşi et al., 2015) or the sufficiently small beaming fraction, which is also expected to play role to some extent in the current statistics of PULXs and non-pulsating ULXs (Erkut et al., 2020).

Next, the neutron star in M51 ULX-8 will be assumed to be close to spin equilibrium. The critical value for the fastness parameter in Equation (10) will therefore be adopted, i.e., $\omega = \omega_c = 0.75$ (Türkoğlu et al., 2017) in addition to the numerical substitution for other quantities such as the source distance $d = 8.58$ Mpc, the 0.5-8 keV flux level $F_X \approx 5.43 \times 10^{-13}$ erg s$^{-1}$ cm$^2$ at which the CRSF was detected (Brightman et al., 2018), the magnetic field $B$ inferred from the observed cyclotron energy at 4.52 keV for both electrons and protons (Equations 14 and 15; Figure 1), the relative width of the boundary region changing from $\delta \approx 0.01$ to $\delta \approx 0.3$ (Erkut & Çatmabacak, 2017; Erkut et al., 2020), and the canonical values $M_* = 1.4 \, M_\odot$ and $R_* = 10$ km for the neutron-star mass and radius, respectively.

Equation (10) is used to explore the parameter space of beaming fraction and spin period. The same numerical values for some of the parameters in Equation (10) are also employed in Equations (7) and (9) to set additional constraints on this parameter space to reveal the allowed region where the beaming fraction and neutron-star spin period are determined concurrently. In Figure 2, the variation of the beaming fraction with the spin period of the neutron star is shown in accordance with Equation (10). The two shaded areas correspond to the electron and proton cyclotron interpretations of the absorption line detected at 4.52 keV in the X-ray spectrum of M51 ULX-8. The boundaries of each shaded area are labeled with the upper and lower limits of $\delta$. The allowed regions for the beaming fraction and spin period are determined by the minimum and maximum values of the beaming fraction (Equations 7 and 9) in addition to the boundaries of shaded areas.

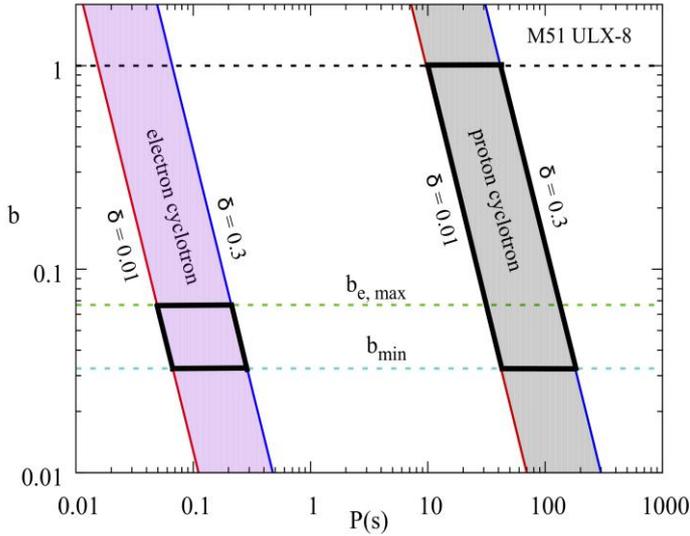

*Figure 2. Beaming fraction as a function of the spin period of the neutron star in M51 ULX-8. The shaded areas bounded by the solid black curves show the allowed regions for the spin period and beaming fraction according to the electron and proton cyclotron interpretations.*

Note from Equation (7) that $b_{min}$ is independent of the magnetic field and therefore is common to both electrons and protons whereas $b_{max}$ depends on the magnetic field (Equation 9). The upper limit of the beaming fraction for electrons $b_{e, max}$ has a much smaller value than the upper limit of the beaming fraction for protons as the magnetic field estimated for the eCRSF is at least three orders of magnitude weaker than the field for the pCRSF (Equations 14 and 15). For the proton cyclotron interpretation, the upper limit of $b$ estimated by the subcritical luminosity condition (Equation 8) exceeds 1 and thus must be replaced by the uppermost limit $b_{max} = 1$, as shown in Figure 2.

In the case of eCRSF, the ranges for the neutron-star spin period and beaming fraction can be deduced from Figure 2 as 0.05 s < $P$ < 0.3 s and 0.03 < $b$ < 0.07, respectively. In a similar way, 10 s < $P$ < 200 s and 0.03 < $b$ < 1 are found for pCRSF.

## 4. Conclusions and Recommendations

Our estimates for the spin period and beaming fraction are based on the canonical values of the neutron-star mass and radius, i.e., $M_* = 1.4\ M_\odot$ and $R_* = 10$ km. The range for the beaming fraction slightly shifts toward higher values by a factor of ~ 1.4 if a higher mass value such as $M_* = 2\ M_\odot$ is preferred for the neutron star, which would yield 0.04 < $b$ < 0.1 for eCRSF and 0.04 < $b$ < 1 for pCRSF (see Equations 7 and 9). The effect of the magnetic field variation on $b_{max}$ when the compactness changes from 1.4 to 2 is negligible (see Figure 1). For a given value of the beaming fraction, the spin period depends on the neutron-star mass and radius as

$$P \propto R_*^{15/7} M_*^{-2/7} \qquad (16)$$

according to Equation (10). Note that $P$ is highly sensitive to the radius. Increasing the neutron-star radius from 10 km to 13 km would raise the spin period by a factor of ~ 1.8.

Even though the CRSF detected at ~ 4.5 keV in the Chandra X-ray spectrum of M51 ULX-8 strongly indicates that the accreting object is a neutron star rather than a black hole, no pulsations could be detected despite long exposure times, most probably because of low time resolution (~ 3.2 s) of the ACIS detectors on board Chandra (Brightman et al., 2018). If true, the allowed region determined in Figure 2 using the electron cyclotron interpretation becomes more likely for M51 ULX-8 to estimate $P$ in the 0.05-0.3 s range and $b$ in the 0.03-0.07 range with a magnetic field strength of ~ $5 \times 10^{11}$ G on the surface of the neutron star for a compactness of 1.4 (Figure 1). The elusiveness of pulsations may then be attributed to the relatively small values of the beaming fraction and spin period, which would otherwise be detected by the forthcoming missions such as Athena carrying the X-IFU detector with a high time resolution of 10 microseconds.

## 5. Acknowledge